\documentclass[preprint,english,aps]{revtex4}
\usepackage[T1]{fontenc}
\usepackage[latin9]{inputenc}
\usepackage{verbatim}
\usepackage{graphics,graphicx,amsmath,amssymb}
\usepackage{graphicx,amsmath,amssymb}
\usepackage[small]{caption}
\usepackage[usenames, dvipsnames]{color}

\makeatletter
\@ifundefined{textcolor}{}
{%
 \definecolor{BLACK}{gray}{0}
 \definecolor{WHITE}{gray}{1}
 \definecolor{RED}{rgb}{1,0,0}
 \definecolor{GREEN}{rgb}{0,1,0}
 \definecolor{BLUE}{rgb}{0,0,1}
 \definecolor{CYAN}{cmyk}{1,0,0,0}
 \definecolor{MAGENTA}{cmyk}{0,1,0,0}
 \definecolor{YELLOW}{cmyk}{0,0,1,0}
 }
\makeatother

\usepackage[ pdftex, plainpages = false, pdfpagelabels, 
                 pdfpagelayout = useoutlines,
                 bookmarks,
                 bookmarksopen = true,
                 bookmarksnumbered = true,
                 breaklinks = true,
                 linktocpage=all,
                 pagebackref=false,
                 colorlinks = true,
                 linkcolor = BrickRed,
                 urlcolor  = blue,
                 citecolor = BrickRed,
                 anchorcolor = green,
                 hyperindex = true,
                 hyperfigures
                 ]{hyperref} 

\usepackage{babel}
\begin{document}
\title{\href{http://necsi.edu/research/social/yemen/} {Conflict in Yemen: From Ethnic Fighting to Food Riots}}
\date{July 24, 2012}  
\author{Andreas Gros, Alexander Gard-Murray and \href{http://necsi.edu/faculty/bar-yam.html}{
Yaneer Bar-Yam}}
\affiliation{
\href{http://www.necsi.edu}{New England Complex Systems Institute} \\ 
238 Main St. Suite 319 Cambridge MA 02142, USA \vspace{2ex}
}


\begin{abstract}
Yemen is considered a global terrorist base for Al-Qaeda and in recent years rampant violence is threatening
social order. Here we show that the socio-economic origins of violence
recently changed. Prior to 2008, violence can be attributed to inter-group conflict between 
ethnically and religiously distinct groups. Starting in 2008, 
increasing global food prices triggered a new wave of violence 
that spread to the
endemically poor southern region with demands for government change and economic concessions. This
violence shares its origins with many other food riots and the more recent Arab Spring. 
The loss of social order and the opportunities for terror organizations can be best addressed by directly
eliminating the causes of violence. Inter-group violence can be addressed by delineating within-country 
provinces for local autonomy of ethnic and religious groups. The impact of food prices can be alleviated 
by direct food price interventions, or by addressing the root causes of global food price increases
in US policies that have promoted conversion of corn to ethanol and commodity speculation. 
Addressing the food prices is the most urgent concern as a new bubble in food prices has been
projected to begin before the end of 2012. 
\end{abstract} 
\maketitle

Violence has been common in Yemen since the founding of a modern state in the southern 
Arabian peninsula nearly 100 years ago \cite{Dresch1989,Dresch2000}, but recent attacks
and social disruption are particularly severe  \cite{Fattah2011,Bernin2009,Boucek2009}.
Yemen is one of the global bases of Al-Qaeda \cite{Katz2003,NYtimesAqap,CFLAQAP,Phillips2007,PradosSharp2007}, with the attacks on the USS Cole in 2000 and American Embassy in 2008 as the most well known local incidents.
Yemeni Al-Qaeda has also been involved in global terror activities including an
alleged attempt to bomb a Detroit-bound plane on Christmas Day of 2009. At the beginning 
of that year, Yemeni Al-Queda joined with the smaller Saudi Al-Queda to form Al Queda of
the Arabian Peninsula (AQAP). The current social disruption increases concerns that Yemen may
become an even stronger terrorist base, threatening security worldwide. 

Here we show that the nature of violence in Yemen has changed between 2005 and
2011 from being ethnically and religiously based to being dominated by the effects of
increases in food prices on an impoverished population. During the period of ethnic and 
religious inter-group violence, geographical locations of incidents are consistent with a theory
that predicts areas of violence based upon the geographical composition of
the population \cite{Lim2007, Rutherford2011}, building on a tradition
of geographic analysis in social science \cite{Berry1968,Chorley1967,Harvey1969}.  
In contrast, the later period of violence begins at the time of globally increasing food prices in
2007, and spread from areas of ethnic conflict in the north to the endemically poor southern part of Yemen.

Our results have immediate implications for strategies to reduce violence and
limit the growth of terrorist influence. Rather than direct military and
security operations, effective interventions may require eliminating the
primary economic and social drivers of violence. First, the immediate economic
drivers can be relieved by addressing the problems of the global food market,
which has been implicated more broadly in the revolutions in North
Africa and the Middle East \cite{Lagi2011,Ciezadlo2011,Brown2011}. Second, by providing partial internal
autonomy to ethnic and religious groups, the origins of the longer term inter-group
violence can also be alleviated \cite{Rutherford2011}.  

The interventions we identify would not only address the growing security
risks, but also improve the living conditions of millions of people,
reducing severe poverty, social disruption and endemic violence. Indeed,
these two goals are directly linked as the social and economic conditions
are the origins of social disorder, in whose shadow terrorist activities can
grow. 

Our analysis of violence in Yemen begins from an understanding of the role of geography 
in conflict between distinct self-identifying groups defined by properties like ancestry, 
culture, language, and religion \cite{Lim2007,Rutherford2011}. In this paper we  
use the term ethnic violence to describe this kind of inter-group conflict. Such
violence is typically though not exclusively directed against or by civilians. When
self-identifying groups are either sufficiently separated or sufficiently well-mixed,
violence is unlikely. Separation limits inter-group friction, while integration inhibits 
inter-group alienation. Ethnic violence occurs most frequently in areas that
have a certain intermediate degree of population separation, but in which 
control of the area is not separated accordingly, i.e. neither political nor
physical boundaries exist to allow for local autonomy. 
In places where self-identifying groups separate into geographical patches of a critical size, in the range of 20-60 km, a group is able to impose its cultural norms, religious values, language differences and in-group social signaling within public spaces. These spaces may include public squares, markets, restaurants, places of worship and schools. However, when social expectations are violated because of the proximity of other ethnic domains, the resulting friction is likely to cause radicalization of some members of the population. Even a small radicalized minority is enough to lead to endemic conflict, and the propensity for violence becomes high. The violence may engage political and military components. Still, the origin of the conflict in the self-identity of the groups is likely to be manifest in violence directed against those who are not politically or militarily powerful. For patches larger than the critical geographical size individuals remain largely within
their own domains and {\it de facto} local sovereignty exists.  If patches are
smaller than the critical size, ethnic groups cannot impose their own norms
and expectations about behavior in public spaces, allowing for the peaceful coexistence of the multiple
ethnic groups that are present. Natural and political boundaries can
increase autonomy to allow for separation that can prevent violence in
areas where it would otherwise occur.  Tests of ethnic violence in various
parts of the world have indeed shown that ethnic violence occurs in the
vicinity of patches of a critical size without well-defined boundaries
\cite{Lim2007, Rutherford2011}.

In contrast to ethnic violence, social unrest reflecting socio-economic despair is
often directed against authorities that fail to satisfy the most basic needs of
the population, especially available or affordable food. Indeed, the relationship between food
prices and social unrest has been 
demonstrated \cite{Lagi2011,ArezkiBruckner2011,Bellemare2011,Bush2010,WaltonSeddon1994}.  Food riots
around the world in 2007-8 and 2010-11 were triggered by steep increases in food
prices. Since Yemen is one of the poorest countries in the Arab World
\cite{CIAFactbookYemen2011}, increases in food prices severely impact a large portion of
the population \cite{IRIN2008}. According to the World Bank's 2007 Poverty Assessment
Report, 35\% of the country's population is classified as poor
\cite{WorldbankPovertyAssessment2007}.

\begin{figure}[tb]
\includegraphics[width=.9\textwidth]{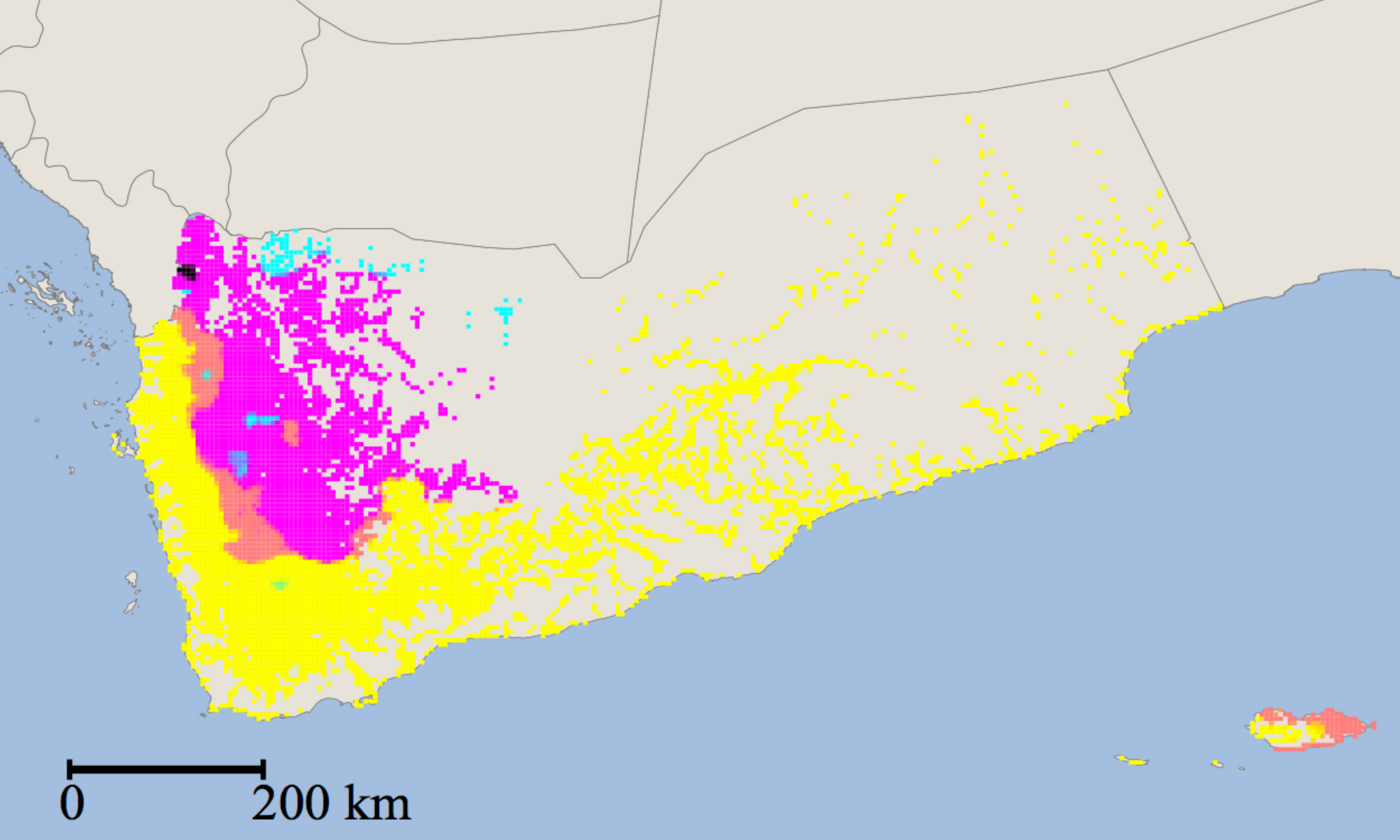}
\caption{Estimated spatial distribution of major self-identifying groups in Yemen in 2000 \cite{Gulf2000}; Zaydi Shiite: magenta, mainstream Sunni: yellow, Ismaili: cyan, Salafi: black
\label{fig:ethnicities}}
\end{figure}

In order to perform a more detailed and quantitative analysis we start by
considering the ethnic geography of Yemen. 
There are four commonly described self-identifying ethnic and religious groups
in Yemen: Zaydi Shiites, Ismaili Shiites, mainstream (Shafi'i) Sunnis and Salafi (Wahhabi) Sunnis. 
Together these groups are estimated to represent 99\% of the population
(55\% Sunni and about 44\% Shiite) \cite{Medea2007}.  Yemen's
societal structure has a strong tribal aspect, especially in rural areas
\cite{Manea1998,Dresch1989}.  While neither tribal allegiances nor political attitudes
necessarily align with their members' religious denomination \cite{Katz2003},
it is nevertheless reasonable to assume as a first approximation that conflict arises
between self-identified ethnic and religious groups. 
Obtaining data about the geographical distribution of these groups is difficult as there
is no direct census and the distribution has changed in recent times, especially due to the 
spread of Salafism \cite{Bonnefoy2012}. Moreover, since political and religious affiliations 
may be linked, various movements including the Moslem Brotherhood may have both
political and religious connotations. For our analysis of ethnicity and
violence we use spatial demographic data from 2004 \cite{YemenCensus2004} to
identify the populated areas and 
an approximate map of the spatial distribution of the four major groups in 2000  
obtained from a compilation of sources \cite{Gulf2000}
to identify ethnic 
compositions, as shown in Figure \ref{fig:ethnicities}.
The approximate nature of the available data limits the precision of the calculations
we perform. Demographic dynamics, specifically the spread of Salafism
in recent years changed the sectarian associations across Yemen.  
Our conclusions only depend on very general features of spatial geography, 
specifically the presence of groups of a given geographic size in a region of the country. 
The conclusions are therefore robust to all but very specific localized changes relative to
surrounding areas. This is a strength of our method, especially in application to areas
where data is poor and changes are ongoing. 
Data on violent incidents was obtained from
the Worldwide Incident Tracking System (WITS) \cite{WITS2012} from
which we selected the incidents that involved civilian casualties.

\begin{turnpage}
\begin{figure}[htb]
	\centering
	\begin{minipage}{.3\textheight}
		\centering
		\includegraphics[width=\textwidth]{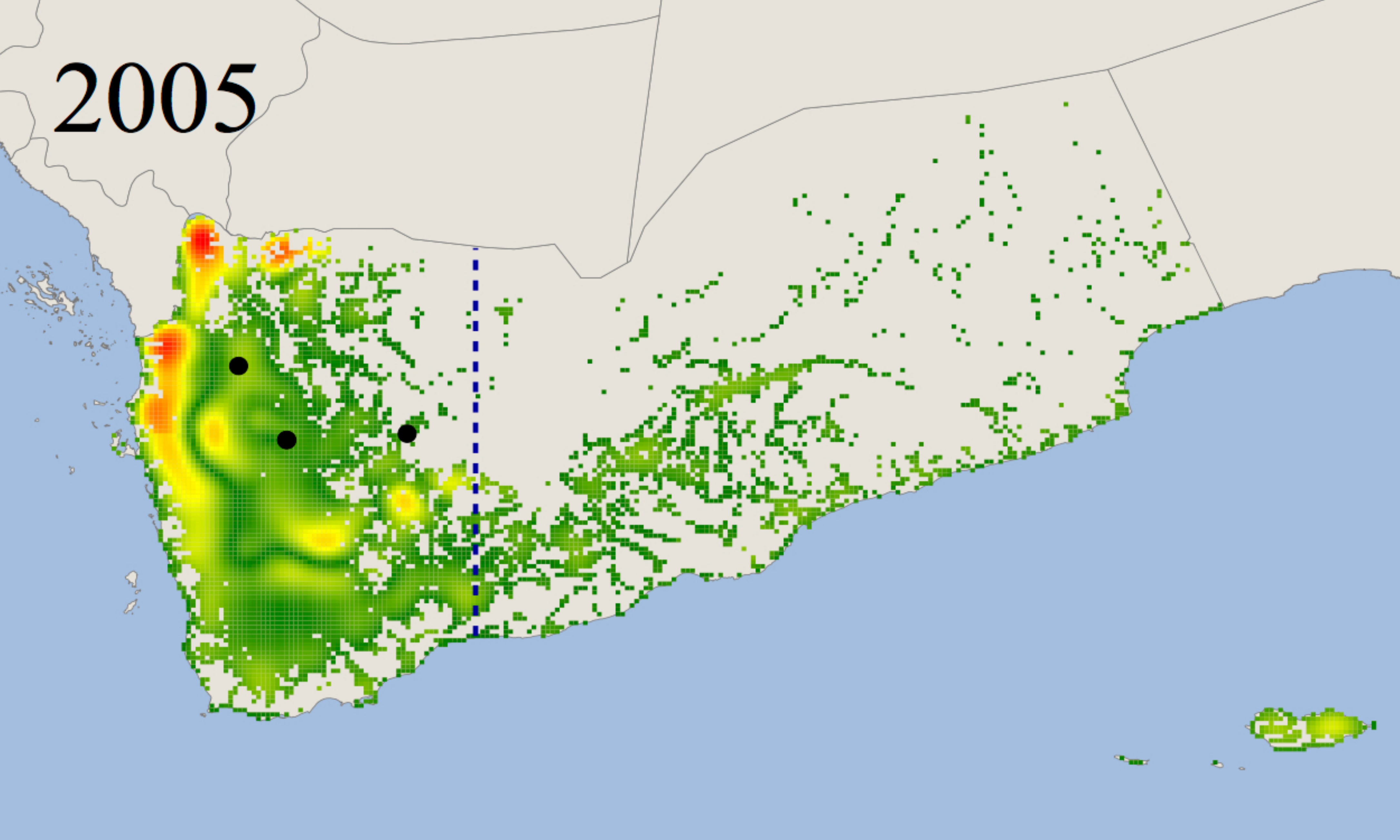}
	\end{minipage} 
	\begin{minipage}{.3\textheight}
		\centering
		\includegraphics[width=\textwidth]{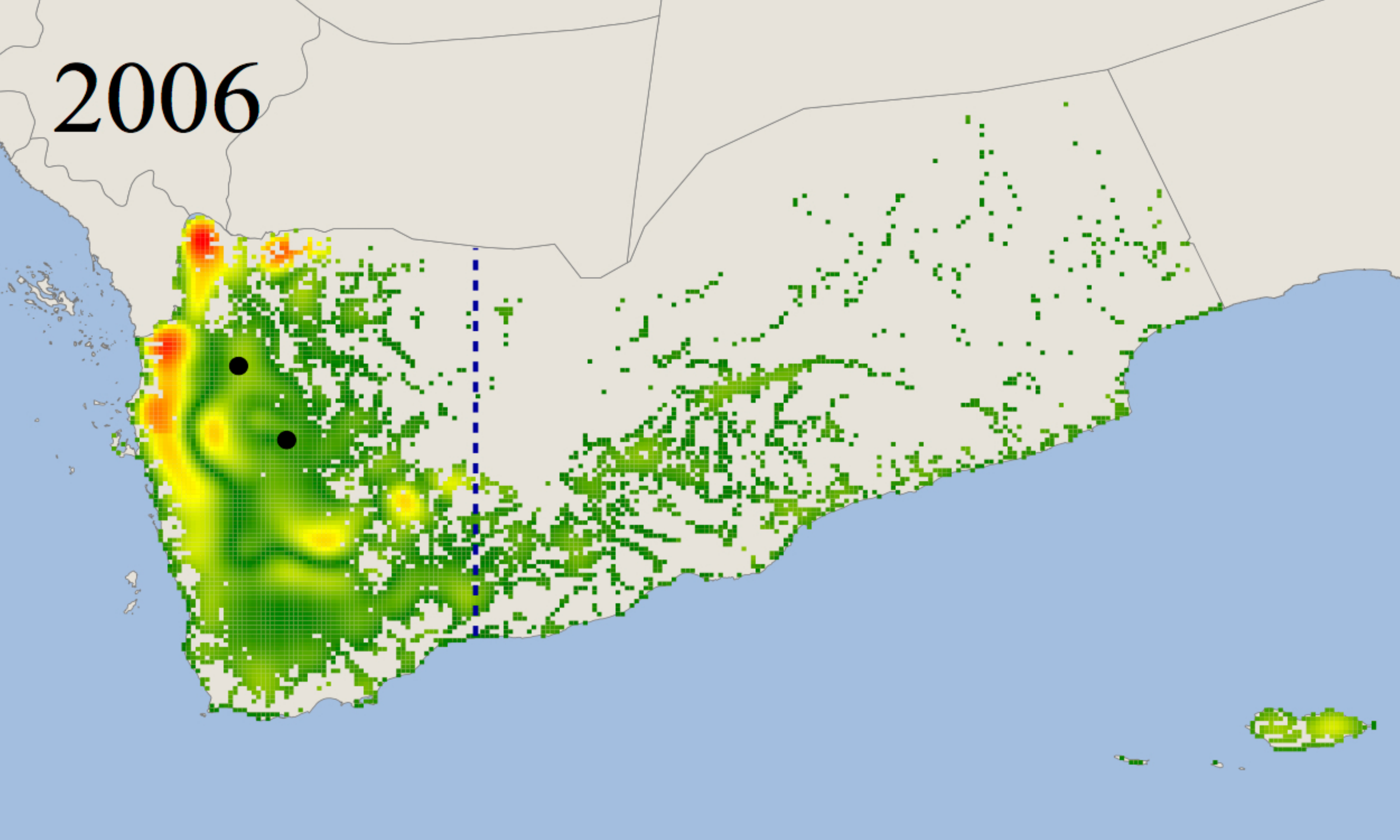}
	\end{minipage} 
	\begin{minipage}{.3\textheight}
		\centering
		\includegraphics[width=\textwidth]{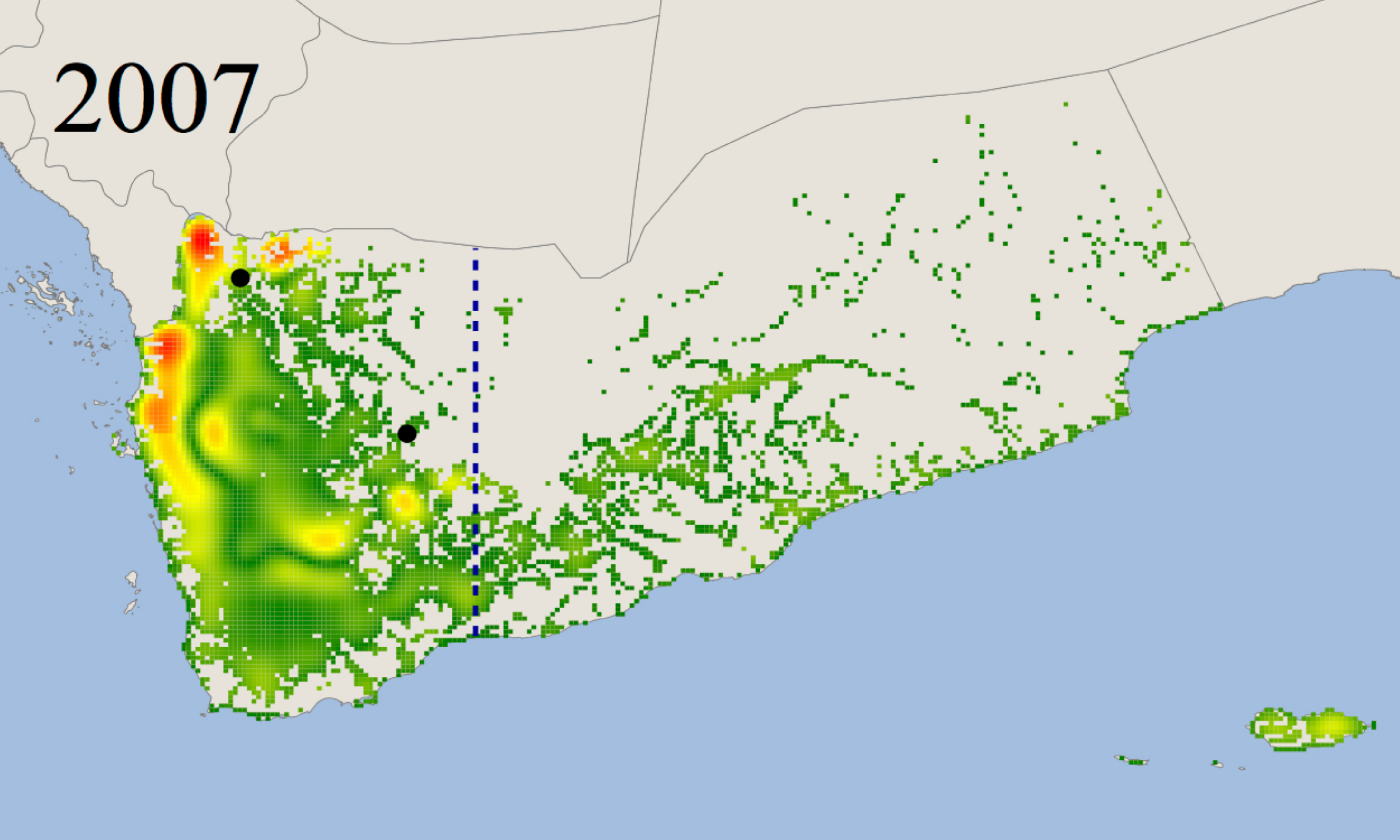}
	\end{minipage} \hspace{\fill}

	\begin{minipage}{.3\textheight}
		\centering
		\includegraphics[width=\textwidth]{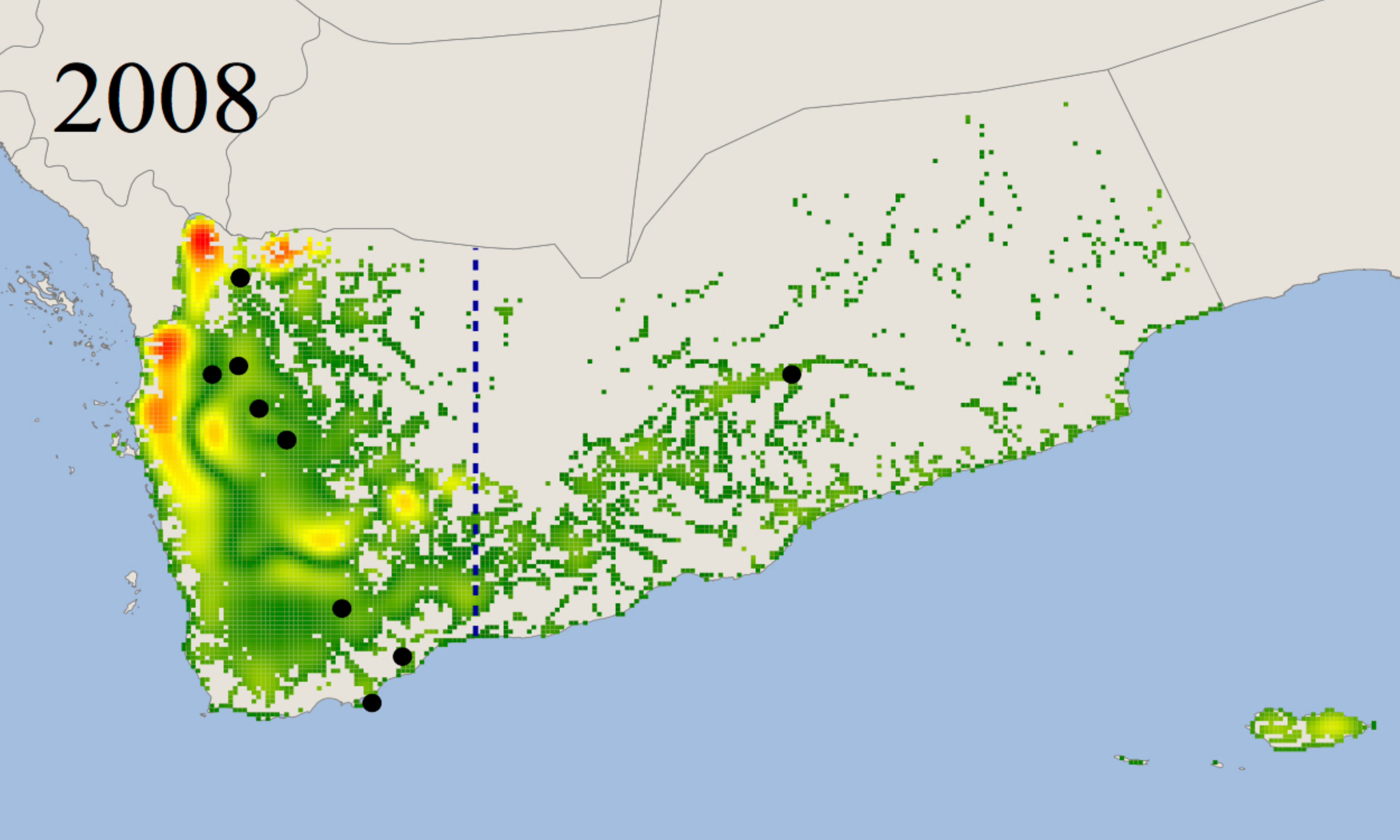}
	\end{minipage} 
	\begin{minipage}{.3\textheight}
		\centering
		\includegraphics[width=\textwidth]{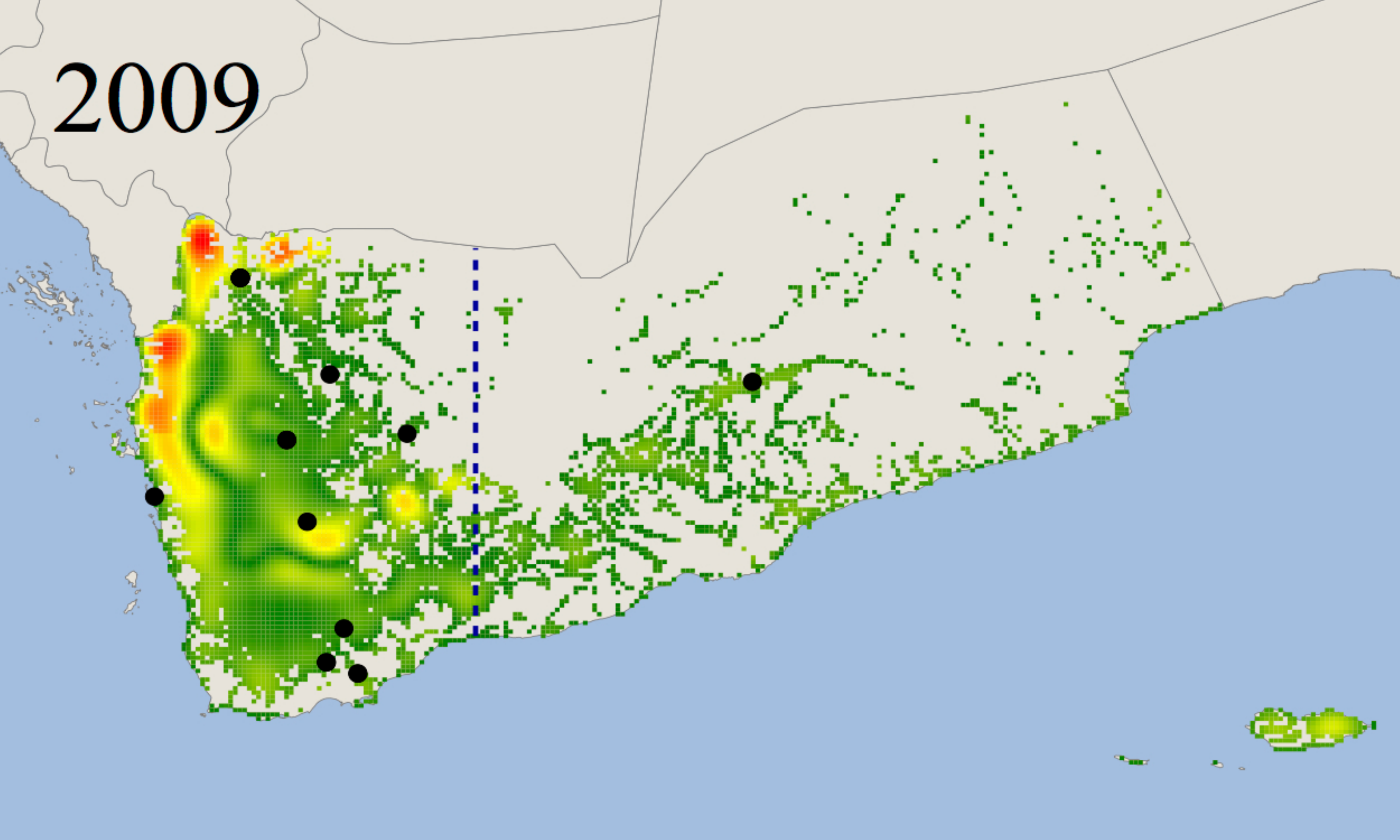}
	\end{minipage} 
	\begin{minipage}{.3\textheight}
		\centering
		\includegraphics[width=\textwidth]{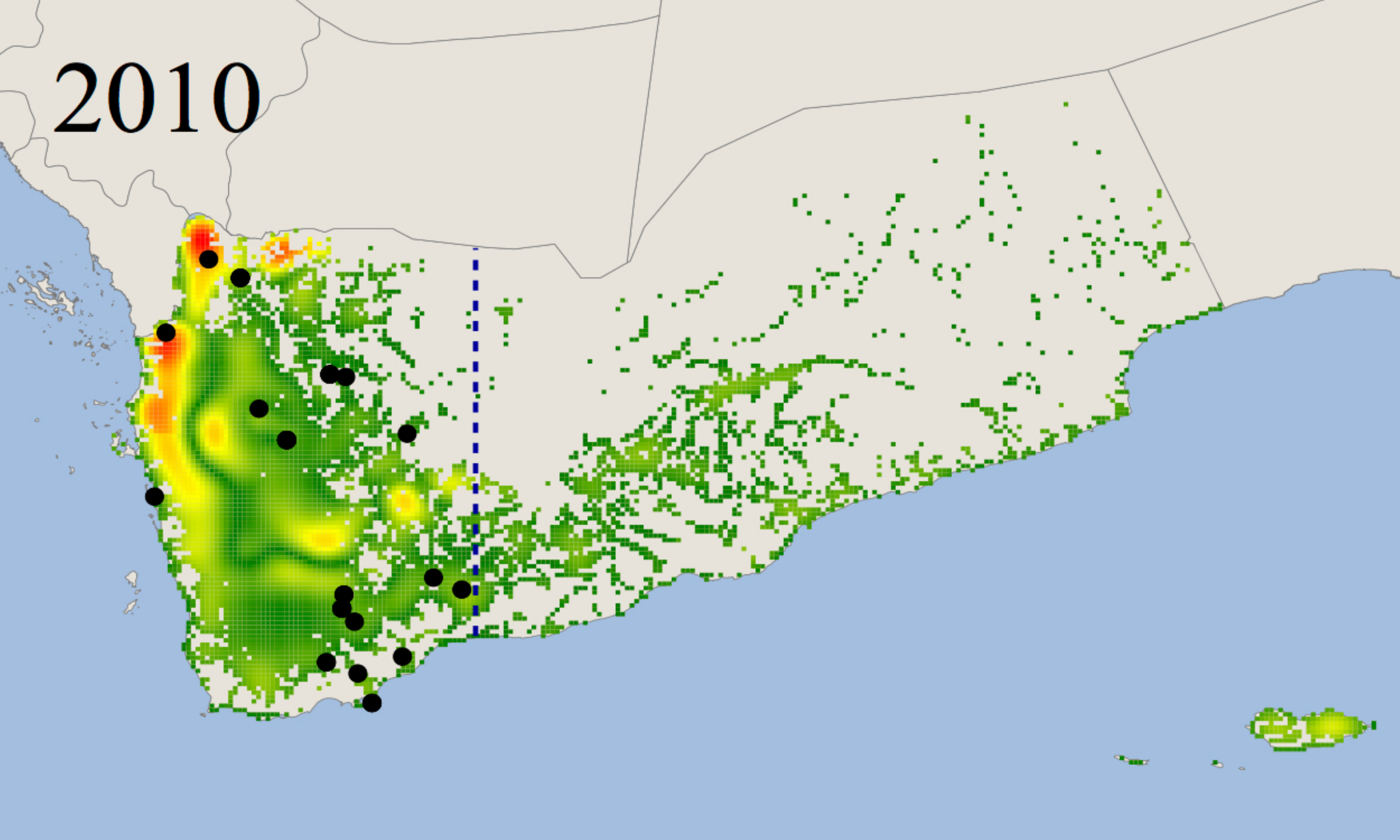}
	\end{minipage}\hspace{\fill}

	\begin{minipage}{.3\textheight}
		\centering
		\includegraphics[width=\textwidth]{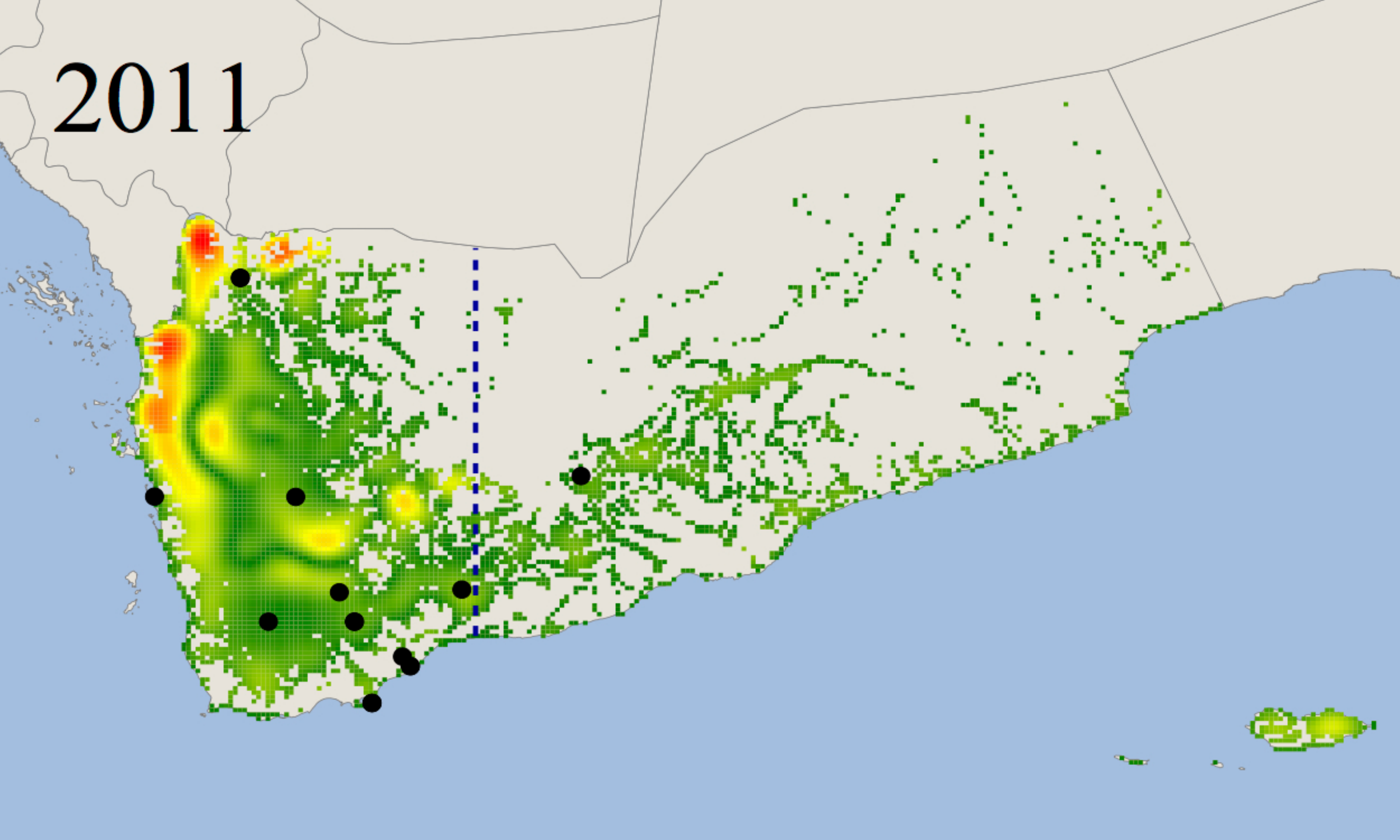}
	\end{minipage} 
	\begin{minipage}{.3\textheight}
		\centering
		\includegraphics[width=\textwidth]{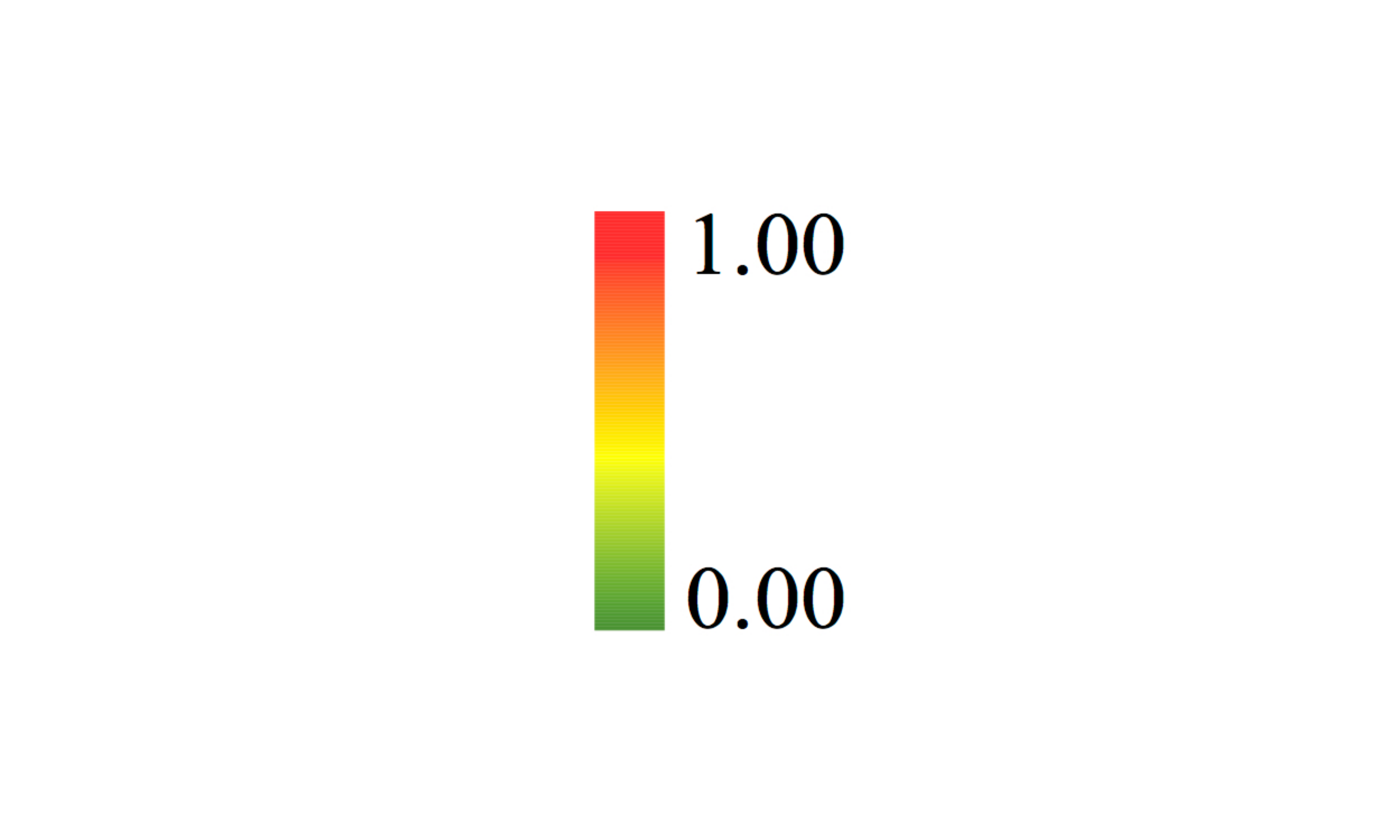}
	\end{minipage} 
	\begin{minipage}{.3\textheight}
		\centering
		\includegraphics[width=\textwidth]{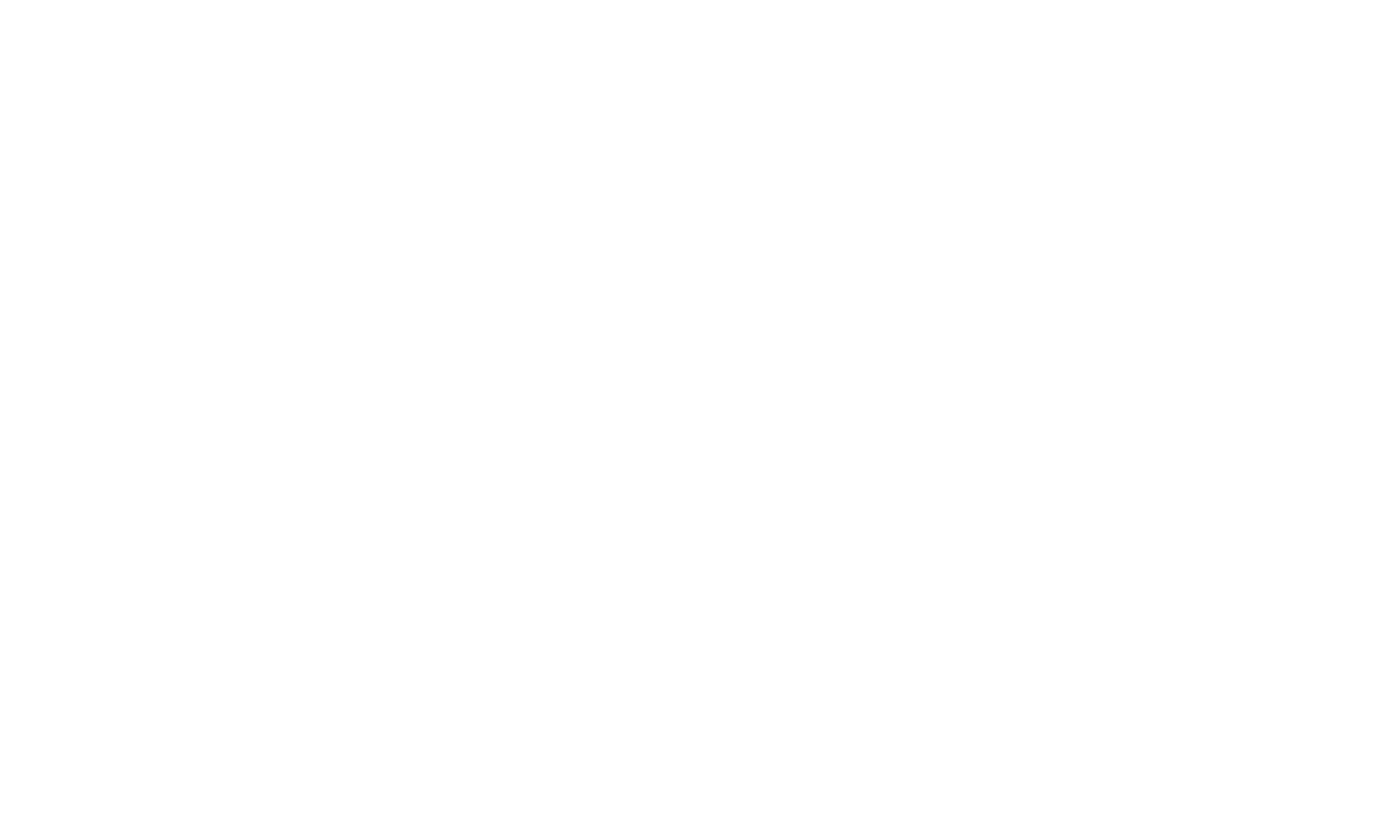}	
	\end{minipage}\hspace{\fill}
	\caption{Propensity for violence (color bar) and incidents of violence (black dots) in populated areas of Yemen. Dashed vertical line delineates the western part of Yemen that we consider in the correlation analysis (Figure \ref{fig:correlations}). Much of the area to the east has a population density of less than 1 person per square km. 
		\label{fig:violence}}
\end{figure}
\end{turnpage}

We calculated the propensity for violence in any given populated area by
identifying patches of ethnic groups of a critical size of 56 km. This size is
consistent with the value that provides predictive success in other countries
\cite{Lim2007, Rutherford2011}. Mathematically we use a wavelet filter
\cite{Lim2007} that weighs the presence of ethnic types in a circular area
around a focal point against the presence of ethnic types in the surrounding
area. If ethnic types are well mixed or the whole area is populated uniformly
by only one type, the output of the filter is small.  However, if the inner area
is populated by a different type than the surrounding area, 
forming an ethnic island or peninsula, the output will be
high.  We perform this analysis for focal points on a fine regular mesh throughout Yemen with
results shown in Figure \ref{fig:violence}. Actual incidents of violence
involving civilians are indicated for each year from 2005 through 2011. 
\clearpage

\clearpage

We quantify the level of agreement between our prediction and the data on
violent incidents by correlating maps of shortest distances to locations of
violent incidents and locations of predicted violence. We calculate the
distance to the closest violent incident and the closest location of predicted
violence at every point on the spatial mesh for a given year.  We consider a
location of predicted violence to be any point where the violence potential is
above a threshold of 0.48 (the average propensity to violence plus two standard
deviations). 

\begin{figure}[tb]
\centering
\includegraphics[width=0.8\textwidth]{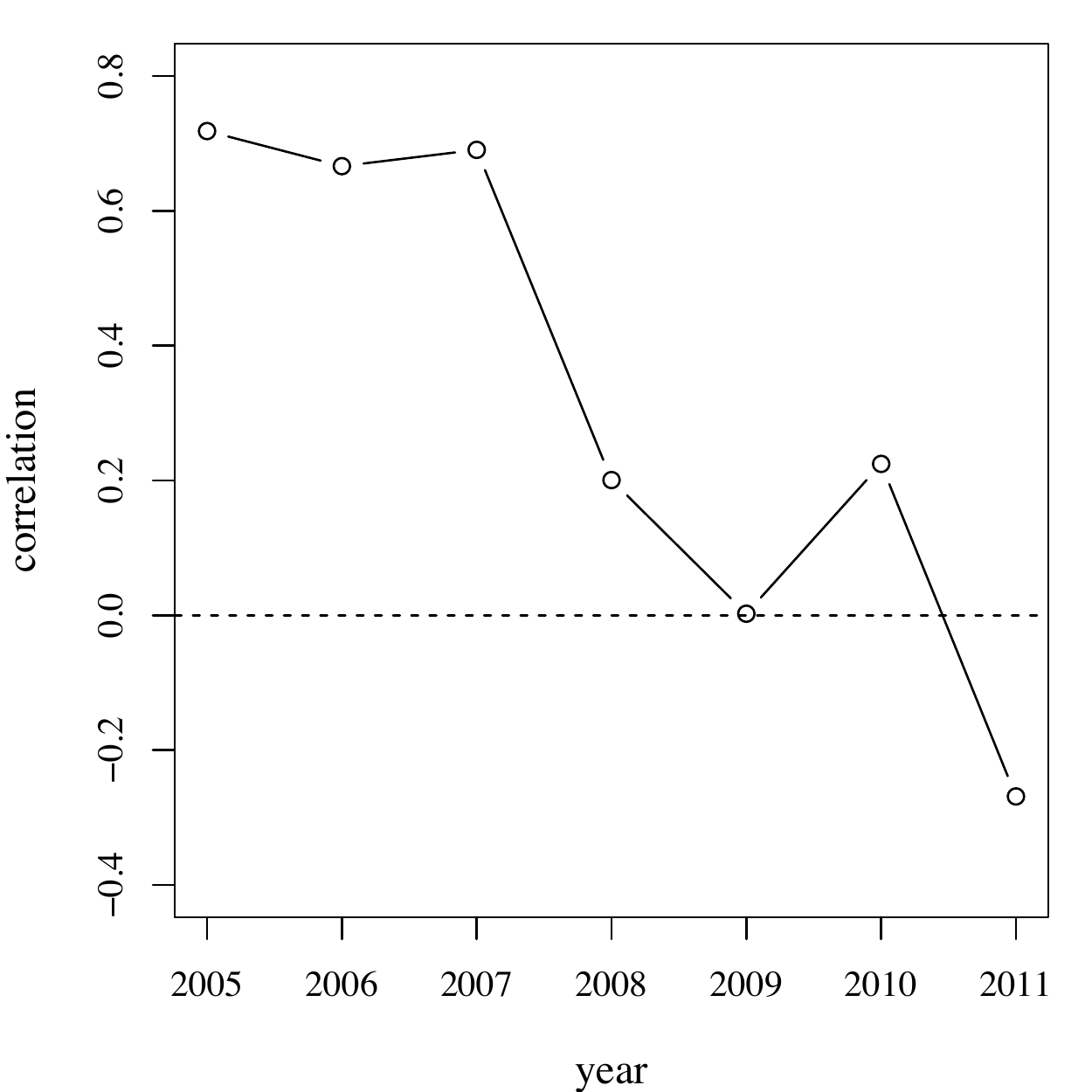}
\caption{Correlations between shortest distances to locations of predicted and actual violence; the
confidence values for the years 2005, 2006, and 2007 are 98.00\%, 96.58\%, and 99.23\%, respectively. 
\label{fig:correlations}}
\end{figure}

We performed the analysis for the western part of Yemen, in which most 
of Yemen's population resides (see vertical dashed line in Figure \ref{fig:violence}).
Figure \ref{fig:correlations} shows the correlations between shortest distances 
to actual and predicted locations of ethnic violence in the west over time. 
The correlation values are approximately 0.70 for 2005, 2006, and 2007, 
and drop to between -0.2 and 0.20 for 2008 through 2011, showing a distinct shift 
away from ethnically-motivated conflict. The confidence values for the
correlations in years 2005, 2006, and 2007 are 98\%, 97\%, and 99\%, respectively. 
We calculate confidence intervals using 100,000
trials with random placement of the same number of predicted locations of
violence within the western part of Yemen and compare the
correlations between the corresponding maps of shortest distances.  
The correlation values for the years 2005, 2006, and 2007 are lower than
reported in previous studies \cite{Lim2007, Rutherford2011}, perhaps due 
to the limitations of the geographic ethnic data and reporting of incidents in the Worldwide
Incident Tracking System for these years.  However, the 
confidence values for 2005, 2006, and 2007 are still well above 95\%.
Our results are consistent with reports that ethnic violence plays a 
significant role in Yemen resulting in the deaths of more 
than 2,000 people annually \cite{IRIN2006}. Some violence is
politicized in the form of the Houthi rebellion, but it also has 
sectarian roots and manifests in violence against civilians
in a manner characteristic of ethnic conflict \cite{IRIN2010,IRIN2012,PakistanToday2011}. 

\begin{figure}[tb]
\centering
\includegraphics[width=0.9 \textwidth]{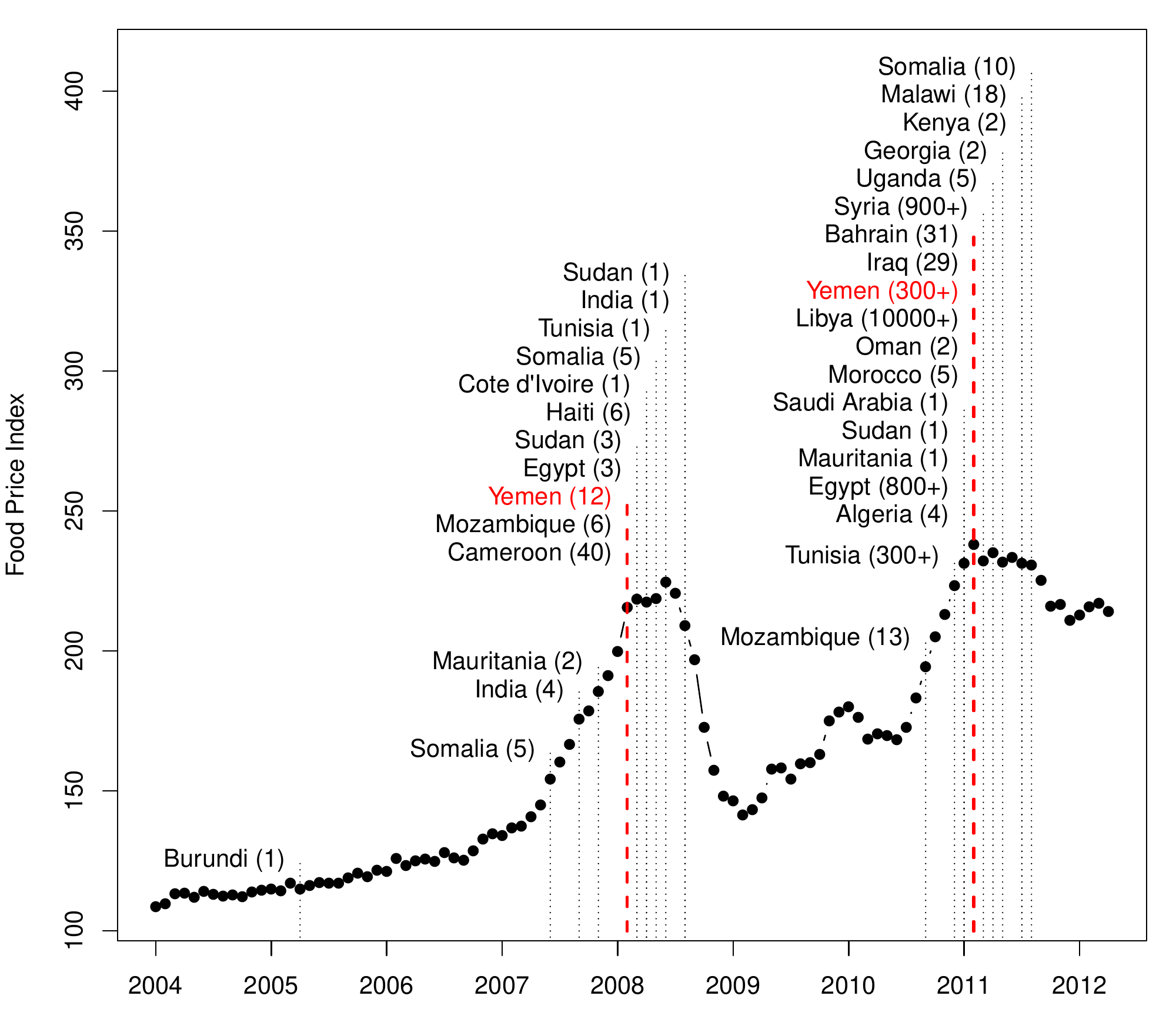}
\caption{Global food price index and the occurrence of food riots (number of casualties in brackets); food riots in Yemen are marked red; reproduced with permission from \cite{Lagi2011} 
\label{fig:foodprices}}
\end{figure}

The marked drop in correlations after 2007 indicates that the nature of the
conflict changed and was no longer solely ethnically motivated. In order to identify
the origins of violence after 2007 we turn to an understanding of social unrest
in which food prices are a key component \cite{Lagi2011}. 
Figure \ref{fig:foodprices} shows the global Food Price Index
over time and the occurrence of food riots and revolutions associated with the
Arab Spring \cite{Lagi2011}. The dates of the food riots in Yemen in early 2008
and 2011 are marked in red and coincide with similar events in many other
countries.  The co-occurrence of global food riots with large spikes in food
prices is consistent with a causal role of food prices in social unrest.
(An alternative hypothesis positing that the spread of Salafism caused the 
violence in the south is not supported by direct analysis, indicating they
were not particularly involved \cite{Bonnefoy2012}.)
Figure \ref{fig:foodprices} shows that food riots in 2007-08 and 2010-11 were
not a local phenomenon, but affected a broad spectrum of regions in Africa and
the Middle East as well as Haiti and India. We can therefore understand the
appearance of social unrest at these times based upon a hypothesis that
widespread unrest is not necessarily related to governmental activities, or,
in the case of Yemen, to terrorist actions. Instead, social unrest is
induced by the government's perceived failure to provide food
security to the population \cite{Lagi2011}. The poverty prevalent in Yemen
\cite{Fattah2011}, and southern Yemen's dependency on imported wheat
\cite{Katz2003,source_grains,IRIN2008}, similar to Egypt and Tunisia \cite{FAOGIEWS}, in combination with rising
food prices, are very likely to have been the underlying trigger for
violent incidents in 2008 and later.

Geographically, the violence in 2008 expanded from the north to the south. 
The southern violence can be understood from the recent political and economic 
history of Yemen.  From 1967 through 1990 South Yemen existed as an independent 
state. The separation between North and South Yemen is partially, but not completely, according to ethnic 
regions. After unification in 1990, the north dominated and the south was economically 
marginalized. Ownership of resources was transferred to northern individuals and 
organizations \cite{Globalsecurity2000, Whitaker2009,SmallArmsSurvey}. The corresponding political
disaffection manifested in a brief civil war in 1994. 
In 2007, during the first food price peak, political discontent coalesced 
into the Southern Movement, which was reenergized by food riots in 2010 to
demand a wide range of economic and social concessions \cite{Time2011,
Day2010}.  While the expansion of violence to the south is a key change,
poverty is also widespread in the north. An increase in violent
activity in 2008 and 2010 in the north can be attributed to 
food based riots overlaid upon the preexisting ethnic conflict. 
Similar to other countries associated with recent revolutions in 
North Africa and the Middle East, the unrest based on food riots developed
into a broader revolutionary process based upon persistent economic 
and social conditions, with implications for both local political instability 
and global terror. 

Our work has identified two major sources of violence in Yemen: partial ethnic separation with
poorly defined boundaries and unreliable food security for a vulnerable
population. These conclusions have direct implications for policy. 

The most urgent socioeconomic problem driving violence is high food prices. 
Recent work has shown that there is likely to be another food price bubble by 
the end of 2012 \cite{Lagi2012}.
Based on this prediction conditions in Yemen will deteriorate if no mediating
policy changes take effect to lower food prices.  Current political efforts to
broaden the governmental basis through assembling a National Dialogue Conference
aim to tackle political grievances but do not address the problems of food prices.
The most direct method to achieve food price stability is to provide subsidies as
have been implemented in many countries in the face of the inability of the population
to afford available food.
Such subsidies are, however, difficult to afford for impoverished countries and would require 
external financing. More fundamentally, while many different factors have been
considered for causing the rise of global food prices, a quantitative analysis has shown
that the drivers of food price increases originate in US agricultural policies that are 
affecting food prices globally. These include two distinct domains
of domestic policy. The first is subsidies for corn-to-ethanol conversion, which resulted 
in growth over less than a decade from negligible rates to 40\% of the US corn crop 
being converted to ethanol \cite{fas2011}. More recently, concerns about their impact has led to
the elimination of these subsidies as of December, 2011 \cite{Pear2012}. 
However, regulations that specify the amount of ethanol to be produced continue \cite{USEnergyIA2007}
The second is the elimination of constraints on 
commodity speculation in 2000 \cite{act2000}, which led to rapid growth of speculative activity through 
commodity index funds that do not follow supply and demand, and result in speculative
bubbles \cite{Kaufman2010,Ghosh2011}. The Commodity Future Trading Commission is in the process of reimposing 
constraints on commodity trading to avoid speculative bubbles \cite{fedreg2011,PattersonTrindle2011}. 
However, the market participants are seeking to dilute the impact of these new regulations \cite{Trindle2012, Brush2012,FIA,Donahue2010,Damgard2011}
These examples show that increased attention to the impact of food prices and their role in global social 
unrest necessarily links global security planning and domestic agricultural policy.

The violence that is ethnic in nature could be dramatically reduced by increasing political
independence by establishing internal country boundaries between ethnic groups \cite{Rutherford2011}. 
The paradigmatic example of the use of internal political boundaries to successfully promote 
peace is that of the Cantons in Switzerland which were established to separate Catholic and 
Protestant populations at a time when conflict was prevalent. The success of this approach
of internal autonomous regions can be considered a model for other areas of the world. 
The value of a federal system of governance to reduce the propensity for violence in Yemen 
has been recently suggested \cite{SEN2011}.
More political self-determination has been demanded by the Southern Movement \cite{Day2010} 
and would most likely be easier to implement than separate nations. One form of potential boundary is the implementation of road
blocks, which are currently used by the government as well as by tribes \cite{Dawood2012}, Al-Qaeda, and 
Ansar al-Sharia \cite{Qaed2012}.  However, access control is met with hostility
where the authority over group territories is not legitimized or established historically.    
Legitimizing partial autonomy in a context of central government power, in regions determined
by the geographical distribution of the main ethnic groups, would be effective according
to our analysis.  

We have shown that science can directly analyze social disruption and violence
and identify their causes, as well as provide recommendations about policy
changes to mitigate them.  Our framework enables us to consider violence within
its socioeconomic context.  Terrorist organizations proliferate in the power
vacuum in countries in which the government is inherently unable to provide
order. For the specific case of Yemen, insurgents benefit from and amplify existing social
disruption as the government and military are caught up in conflicts stemming
from food insecurity and ethnic differences. Food prices and ethnic conflict can
be seen to play a direct role \cite{Swift2012}. We recommend the implementation of
jointly defined internal political boundaries, within which the different
groups can enjoy a degree of self-determination, in addition to lowering food
prices through within-country subsidies and global food policy actions, as
these measures have the greatest chance of stabilizing Yemen.  

We thank Jeb Boone and Charles Schmitz for helpful comments on the manuscript. 
This work was supported in part by AFOSR under grant FA9550-09-1-0324.

\clearpage
\newpage


\end{document}